%% file: main.tex
\documentclass[10pt,final]{IEEEtran}

\usepackage{mathtools}
\usepackage{graphicx}
\usepackage{bm}
\graphicspath{{figs/}}
\usepackage[utf8]{inputenc} % allow utf-8 input
\usepackage[T1]{fontenc}    % use 8-bit T1 fonts
\usepackage[dvipsnames]{xcolor}         % colors
\usepackage[noadjust]{cite}
\usepackage{hyperref}
\hypersetup{hidelinks}
\usepackage{url}            % simple URL typesetting
\usepackage{booktabs}       % professional-quality tables
\usepackage{amsmath,amssymb,amsfonts,amsthm,stmaryrd}
\usepackage{nicefrac}       % compact symbols for 1/2, etc.
\usepackage{microtype}      % microtypography
\usepackage{authblk}
\usepackage[linesnumbered,ruled,vlined]{algorithm2e}
\usepackage{subcaption}
\usepackage{dblfloatfix}
\usepackage{dsfont}
\usepackage[misc]{ifsym} % Add the Letter sign

\usepackage{siunitx}
\usepackage{threeparttable}
\input{C-macro.tex}

\title{Multi-Metric AutoRec for High Dimensional and Sparse User Behavior Data Prediction}

\author{
    \IEEEauthorblockN{
      Cheng~Liang \IEEEauthorrefmark{1},
      Teng~Huang \IEEEauthorrefmark{1},
      Yi~He \IEEEauthorrefmark{2}
      Song~Deng \IEEEauthorrefmark{3}
      Di~Wu \IEEEauthorrefmark{4}
      Xin~Luo \IEEEauthorrefmark{4}
    } 

    \IEEEauthorblockA{\IEEEauthorrefmark{1}
        Institute of Artificial Intelligence and Blockchain, Guangzhou University, Guangzhou, China
    }

    \IEEEauthorblockA{\IEEEauthorrefmark{2}
    Department of Computer Science, Old Dominion University, Norfolk, Virginia 23529, USA
    }

    \IEEEauthorblockA{\IEEEauthorrefmark{3}
        Institute of Advanced Technology, Nanjing University Post \& Telecommunication, Nanjing, China
    }

    \IEEEauthorblockA{\IEEEauthorrefmark{4}
     College of Computer and Information Science, Southwest University, Chongqing 400715, China
    }

  \IEEEauthorblockA{
    \IEEEauthorrefmark{1} c\_liang@e.gzhu.edu.cn, huangteng1220@buaa.edu.cn
    \IEEEauthorrefmark{2} yihe@cs.odu.edu
    \IEEEauthorrefmark{3} dengsong@njupt.edu.cn
    \IEEEauthorrefmark{4} wudi.cigit@gmail.com, luoxin21@gmail.com
  }
}

\begin{document}
\onecolumn
\pagenumbering{gobble} % Show/not page numbers
\maketitle

% abstract
\begin{abstract}
User behavior data produced during interaction with massive items in the significant data era are generally heterogeneous and sparse, leaving the recommender system (RS) a large diversity of underlying patterns to excavate. Deep neural network-based models have reached the state-of-the-art benchmark of the RS owing to their fitting capabilities. However, prior works mainly focus on designing an intricate architecture with fixed loss function and regulation. These single-metric models provide limited performance when facing heterogeneous and sparse user behavior data. Motivated by this finding, we propose a multi-metric AutoRec (MMA) based on the representative AutoRec. The idea of the proposed MMA is mainly two-fold: 1) apply different $L_p$-norm on loss function and regularization to form different variant models in different metric spaces, and 2) aggregate these variant models. Thus, the proposed MMA enjoys the multi-metric orientation from a set of dispersed metric spaces, achieving a comprehensive representation of user data. Theoretical studies proved that the proposed MMA could attain performance improvement. The extensive experiment on five real-world datasets proves that MMA can outperform seven other state-of-the-art models in predicting unobserved user behavior data.
\end{abstract}

\input{Introduction}
\input{Related_Work}
\input{The_proposed_MMA_Model}
\input{Experiment}
\input{Conclusion}

\bibliographystyle{IEEEtran}
\bibliography{ref}

\end{document}

%% file: C-macro.tex
\renewcommand{\eqref}[1]{Eq.\,(\ref{#1})}

\newcommand{\yA}[1]{\hat{y}_{\text{obs}}}
\newcommand{\yR}[1]{\hat{y}_{\text{rec}}}

\newcommand{\romup}[1]{\uppercase\expandafter{\romannumeral #1\relax}}
\newcommand{\romlo}[1]{\lowercase\expandafter{\romannumeral #1\relax}}

\newcommand{\rom}[1]{\uppercase\expandafter{\romannumeral #1\relax}}
\newcommand{\romsm}[1]{\lowercase\expandafter{\romannumeral #1\relax}}

%% file: Introduction.tex
\section{Introduction}
\label{sec:introduction}

Matrices are specifications that describe pairwise relationships between entities that have a wide range of application scenarios~\cite{he2020partial, he2020toward}, such as bioinformatics~\cite{cortazar2018cancertool}, industrial manufacturing~\cite{alhayani2021visual}, and recommendation system(RS)~\cite{zhang2019deep, wu2022prediction}, to mention a few. For instance, the user-item rating matrix has been widely adopted to record the interaction between users and items in various systems. Typically, the rows in the matrix represent users, the columns represent items, and the entries record the results of their interactions.

The hinge of an RS analyzing these rating matrices and completing the missing data lies in handling their sparsity and diversity~\cite{luo2018fast, sun2022toward, liu2022adam, chen2022mnl}. These characteristics conceptually and practically exist in the user behavior data. Nowadays, an RS usually has a large number of items, leaving the user behavior data to be generally sparse and incomplete. Moreover, the different types of interaction and systems cause the diversity of user behavior data which needs well-designed methods to excavate the hidden knowledge.

Recently, with the development of deep learning, the deep neural network (DNN) has been wildly adopted to implement an RS\cite{zhang2019deep, deng2022quantitative, li2022using}. Owing to the nonlinearity, many DNN-based RSs have been proposed to achieve state-of-the-art performance. Despite the success of the DNNs-based models, the same essence the prior models share is that the usage of the loss function and regularization of the model is exclusive and fixed, leaving the model with the single-metric representation capability to the objective user behavior data. However, the real-world user behavior data are heterogeneous and diversified and with various underlying properties~\cite{wu2022double,wu2021latent, wu2021l1}, which are well-studied in other applications (e.g., image processing~\cite{wang2018detecting}), manifesting the limitation of the exclusive and fixed usage of the loss function and regularization~\cite{AutoRec, NRR, sparce-fc, wu2020data, wu2020pmlf}.

Motivated by this finding, the central question of this paper explores: \textit{Can DNNs-based models benefit from different metric spaces and outperform prior works in unobserved user behavior data prediction?}

By proposing a Multi-Metric AutoRec (MMA) for unobserved user behavior data prediction, we offer an affirmative answer to this question. The main idea of MMA is two-fold: 1) Different $L_p$-norms ($L_1$-norm and $L_2$-norm) are adopted in loss function and regularization to develop four variants; b) Aggregating these variant models through self-adaptive and self-tuning weighting strategy to form the final model. Through these methods, the proposed MMA enjoys a multi-metric orientation empowering it to outperform state-of-the-art models in completing the missing entities.

This paper has the following contributions:

\begin{itemize}
    \item It proposes the MMA, which tries to make an accurate prediction for missing entries of user behavior data metrics by exploiting multi-metric representation space.
    \item Theoretical research and experiments prove that the proposed MMA can aggregate metric orientating from the base models.
    \item Algorithm design is conducted on the proposed MMA.
    
\end{itemize}

Experimental results on five real-world benchmark datasets demonstrate that MMA significantly outperforms both non-DNN-based and DNN-based single-metric state-of-the-art models in prediction accuracy.

%% file: Related_Work.tex
\section{Related Works}
\label{sec:related works}

\subsection{LFA-based Model}
The LFA-based models are widely adopted to implement an RS~\cite{srifi2020recommender}. However, most of the prior works adopt only one fixed loss function or a fixed regularization to train the model\cite{li2022using, li2022second}, including matrix factorization-based~\cite {MF,FML, wu2019posterior, wu2019deep, wu2019data, he2019online, luo2022momentum, wei2022robust}, dual regularization-based~\cite{wu2018dual}, kalman-filter-based~\cite{yuan2022kalman}, item content-based~\cite{zhang2019bridging}, generalized non-negative and with momentum-based~\cite{luo2018fast}, and covering-based neighborhood-aware model~\cite{zhang2019covering} to name a few. Despite that, the $L_2$ loss function has been proven to be more sensitive to the outliers but more stable, the $L_1$ loss function is more robust to the outliers but unstable, and $L_1$ regularization has the built-in feature selection characterise~\cite{wu2018l1, song2022non, li2022novel}. Hence, the LFA-based recommender system that is has different $L_p$-norms are proposed~\cite{raza2020regularized, wu2020robust, zhu2018similarity, luo2022neulft}. Yet, there is currently a lack of discussions on multi-metric DNN-based RS.
\subsection{Deep Learning-based Model}
Deep learning-based models are wildly used to predict user behavior data~\cite{wang2022multi, jin2022neural, liu2022symmetry, li2022diversified, jin2022distributed} owning to their non-linearity. Some researchers have comprehensively reviewed the most recent DNN-based RS~\cite{zhang2019deep}. Among these DNN-based methods, autoencoder-based is one of the representative methods~\cite{AutoRec} and stems the variational autoencoder-based method~\cite{liang2018variational}, kernelized synaptic-based autoencoder~\cite{sparce-fc} and global and Local Kernels-based autoencoder~\cite{Glocal-k}. And other representative research includes: neural rating regression-based~\cite{NRR}, federated meta-learning-based~\cite{lin2020meta}, and metric learning via memory attention-based~\cite{tay2018latent}. Additionally, graph neural network (GNN) based RS, for example inductive matrix completion-based~\cite{IGMC}, are also proposed for the user behavior data representation. 

Compared with the above methods, MMA has the following significance. Firstly, he nonlinear features of DNN-based models endow them with substantial data representation capabilities. Secondly, the DNN-based models do not require complex graph data, unlike GNN-based models, so the consumption of resources lower. Section~\ref{sec:experiments} presents numerical experiments and comparisons with the state-of-the-art models to demonstrate the performance advantages of the proposed MMA model.

%% file: The_proposed_MMA_Model.tex
\section{The Proposed MMA Model}
\label{sec:MMA}

To combine all advantages, we proposed the MMA, in which the architecture and prediction process is two parts. Firstly, it uses the observed data of $X$ to train four variant models. Then it ensembles these predictions with a self-adaptive strategy to get the final output. The following content explains the establishment of the base models, the weighting method, and the theoretical analysis.

\subsection{Establishment of base Model}
DNN-based model's characteristic originates from its loss function.  We employ different $L_p$-norms in loss function $l(\cdot)$ and regularization $r(\cdot)$ to establish the base models that summarized in Table~\ref{tab:loss_reg}.

\begin{table}[h]
    \centering
    \small
    
    \caption{Summarization of four base variant models}
    \resizebox{0.5\linewidth}{!}{
    \begin{tabular}{cccc}
        \toprule
        \midrule
        \textbf{base models} & \bm{$l(\triangle_{j,k})$} & \bm{$r(w_i)$} & \textbf{Characteristic} \\
        \midrule
        MMA-1 & $|\triangle_{j,k}|$ & ${\|w_i\|}_1$ &Robustness \& Feature Selection \\
        MMA-2 & $|\triangle_{j,k}|$ & ${\|w_i\|}^2_2$ &\begin{tabular}[c]{@{}c@{}}Robustness \& Fine representation\end{tabular} \\
        MMA-3 & ${(\triangle_{j,k})}^2$ & ${\|w_i\|}_1$ &Stability \& Feature Selection \\
        MMA-4 & ${(\triangle_{j,k})}^2$ & ${\|w_i\|}^2_2$ &\begin{tabular}[c]{@{}c@{}}Stability \& Fine representation \end{tabular} \\
        \bottomrule
    \end{tabular}
    }
    \label{tab:loss_reg}
\end{table}

\subsubsection{MMA-1 ($L_1$ as Loss Function and Regularization)}
The objective function of MMA-1 is as follows:
\begin{align}
L(f)= \sum_{\mathbf{x}^{(k)} \in X}&\left\|\left(\mathbf{x}^{(k)}-f\left(\mathbf{x}^{(k)}; \theta\right)\right) \odot \mathbf{m}^{(k)}\right\|_1 \\
&+\frac{\lambda}{2} \cdot\left({\left\|w_{1}\right\|}_1+\cdots+{\left\|w_{I}\right\|}_1\right). \nonumber
\end{align}

\subsubsection{MMA-2 ($L_1$ as Loss Function and $L_2$ as Regularization)}
The objective function of MMA-2 is as follows:
\begin{align}
L(f)= \sum_{\mathbf{x}^{(k)} \in X}&\left\|\left(\mathbf{x}^{(k)}-f\left(\mathbf{x}^{(k)} ; \theta\right)\right) \odot \mathbf{m}^{(k)}\right\|_1 \\
&+\frac{\lambda}{2} \cdot\left({\left\|w_{1}\right\|}^2_2+\cdots+{\left\|w_{I}\right\|}^2_2\right). \nonumber
\end{align}
\subsubsection{MMA-3 ($L_2$ as Loss Function and $L_1$ as Regularization)}
The objective function of MMA-3 is as follows:
\begin{align}
L(f)= \sum_{\mathbf{x}^{(k)} \in X}&{\left\|\left(\mathbf{x}^{(k)}-f\left(\mathbf{x}^{(k)} ; \theta\right)\right) \odot \mathbf{m}^{(k)}\right\|}_2^2 \\
&+\frac{\lambda}{2} \cdot\left({\left\|w_{1}\right\|}_1+\cdots+{\left\|w_{I}\right\|}_1\right). \nonumber
\end{align}

\subsubsection{MMA-4 ($L_2$ as Loss Function and $L_2$ as Regularization)}
The objective function of MMA-4 is as follows:
\begin{align}
L(f)= \sum_{\mathbf{x}^{(k)} \in X}&{\left\|\left(\mathbf{x}^{(k)}-f\left(\mathbf{x}^{(k)} ; \theta\right)\right) \odot \mathbf{m}^{(k)}\right\|}_2^2 \\
&+\frac{\lambda}{2} \cdot\left({\left\|w_{1}\right\|}^2_2+\cdots+{\left\|w_{I}\right\|}^2_2\right). \nonumber
\end{align}

\subsection{Self-Adaptively Ensemble}
Ensemble learning is the perfect method for aggregating multi-models. It requires the base model to be diverse and accurate. The base models of the proposed MMA are built on different $L_p$-norm, guaranteeing the base models' diversity. Furthermore, the representative AutoRec guarantees the accuracy of its base models. Thus, the base models of the proposed MMA satisfy the two requirements. We further adopt the self-adaptively aggregation method to control the weighting of the base models according to their \textit{loss} on the validation set. The idea means that increase the weight of the $t_{th}$ base model if its \textit{loss} decreases in the $n_{th}$ training iteration or reduce its weight otherwise.

\textbf{Definition 1 (Separate \textit{Loss} of Base Models).} We use $Sl^{t}(n)$ to denote the separate \textit{loss} of the $n_{th}$ iteration of $t_{th}$ base model, which is computed as follows:
\begin{align}
    Sl^{t}(n)= &\sqrt{{ \sum_{j \in J, k \in K}\left (\left(x_{j,k} - \hat{x}^t_{j,k} \right) \times m_{j,k} \right )}^2/ {\|X\|}_0} \\ &\hat{x}^t_{j,k} = f^t(j,k;\theta) \text{ s.t.} \ t = 1,2,3,4, \nonumber
\end{align}
where ${\|\cdot\|}_0$ represents the $L_0$-norm of a matrix which indicates the number of non-zero elements.

\textbf{Definition 2 (Accumulative \textbf{Loss} of Base Models).} The accumulative \textit{loss} $Al^t(n)$ of $Sl^t$ until $n_{th}$ iteration is calculated as follows:
\begin{align}
    Al^{t}(n)= \sum_{h=1}^n Sl^t(h).
    \label{eq:al_base_model}
\end{align}

\textbf{Definition 3 (Ensemble Weight of Base Models).} The ensemble weight $\varepsilon^t$ of the $t_{th}$ base model is calculated as follows:
\begin{align}
    \varepsilon^{t}(n)=\frac{e^{-\delta Al^{t}(n)}}{\sum_{t=1}^{4} e^{-\delta Al^{t}(n)}},
    \label{eq:ensembel_weight}
\end{align}
where $\delta$ is the balance factor to control the ensemble weights of aggregation during the training process.
Based on definitions 1-3, the prediction of MMA in $n_{th}$ iteration is represented as follows:
\begin{align}
    \hat{x}_{j, k}=\sum_{t=1}^{4} \varepsilon^{t}(n) \hat{x}_{j, k}^{t}.
    \label{eq:prediction}
\end{align}

%% file: Experiment.tex
\section{Experiments}
\label{sec:experiments}

\begin{table}[!t] 
	\centering
	\small
	\caption{Properties of all the datasets.}
	\resizebox{0.5\linewidth}{!}{
	    \begin{threeparttable}
    	\begin{tabular}{cccccc}
    		\toprule
    		\midrule
    		{\bf No.} & {\bf Name} & {\bf |\emph{M|}} & {\bf \emph{|N|}} & \bm{$H_o$} & {\bf Density\tnote{*}}  \\
    		\midrule
    		D1 &	MovieLens\_1M &	\num{6040} &	\num{3952}	&\num{1000209} &	4.19\% \\
            D2 &	MovieLens\_100k &	\num{943} &	\num{1682} &	\num{100000} &	6.30\% \\
            D3 &	MovieLens\_HetRec &	\num{2113} &	\num{10109} &	\num{855598} &	4.01\% \\
            D4 &	Yahoo &	\num{15400} &	\num{1000} &	\num{365704} & 2.37\% \\
            D5 &	Douban &	\num{3000} &	\num{3000} &	\num{136891} &	1.52\% \\
    		\bottomrule
    		\bottomrule
    	\end{tabular}
    	\begin{tablenotes}
        \footnotesize
        \item[*]Density denotes the percentage of observed entries in the user-item matrix.
      \end{tablenotes}
    \end{threeparttable}
	}
	\label{tab:dataset}
\end{table}

\begin{table}[tbp] 
	\centering
	\small
	\caption{Descriptions of all the contrasting models.}
	\resizebox{\linewidth}{!}{
	\renewcommand\arraystretch{0.6}
	\begin{tabular}{cc}
		\toprule
		\midrule
		{\bf Model} & {\bf Description}\\
		\midrule
		\begin{tabular}[c]{@{}c@{}} MF\\ \cite{MF} \end{tabular} & \begin{tabular}{p{\columnwidth}}
        It is the representative matrix factorization model that factorizes data of user-item matrix for the recommender systems. \emph{Computer 2009.}\end{tabular}\\
        \midrule
        \begin{tabular}[c]{@{}c@{}} AutoRec\\ \cite{AutoRec} \end{tabular} & \begin{tabular}{p{\columnwidth}}
       It is the representative DNN-based model in representing user-item data for the recommender system. \emph{WWW 2015}.\end{tabular}\\
        \midrule
        \begin{tabular}[c]{@{}c@{}} NRR\\ \cite{NRR} \end{tabular} & \begin{tabular}{p{\columnwidth}}
        It is a DNN-based multi-task learning framework for rating prediction in a recommender system. \emph{SIGIR 2017}.\end{tabular}\\
        \midrule
        \begin{tabular}[c]{@{}c@{}} SparceFC\\ \cite{sparce-fc} \end{tabular} & \begin{tabular}{p{\columnwidth}}It is a DNN-based model which reparametrize the weight matrices in low-dimensional vectors to capture important features. \emph{ICML 2018}.\end{tabular}\\
        \midrule
        \begin{tabular}[c]{@{}c@{}} IGMC\\ \cite{IGMC} \end{tabular} & \begin{tabular}{p{\columnwidth}}
        It is a GNN-based model which can inductive matrix completion without using side information. \emph{ICLR 2020}.\end{tabular}\\
        \midrule
        \begin{tabular}[c]{@{}c@{}} FML\\ \cite{FML} \end{tabular} & \begin{tabular}{p{\columnwidth}}
        It is a matrix factorization model that combine metric learning (distance space) and collaborative filtering. \emph{IEEE TII 2020}.\end{tabular}\\
        \midrule
        \begin{tabular}[c]{@{}c@{}} GLocal-K\\ \cite{Glocal-k} \end{tabular} & \begin{tabular}{p{\columnwidth}}It is a DNN-based model that generalize and represent user-item data into a low dimensional space with a small number of important features. \emph{CIKM 2021}.\end{tabular}\\
		
		\bottomrule
	\end{tabular}
	}
	\label{tab:competitor}
\end{table}

\subsection{General Settings}
{\bf Datasets.} Five frequently used benchmark datasets are chosen to conduct the following experiments. They are real datasets from different fields including e-commerce, movie review sites. TABLE~\ref{tab:dataset}. summarizes their details. MovieLens\_1M, MovieLens\_100k and MovieLens\_HetRec are collected from the movie recommendation website MovieLens\footnote{https://grouplens.org/datasets/movielens/}. Yahoo is collected from the Yahoo  website\footnote{https://webscope.sandbox.yahoo.com/catalog.php?datatype=r}. Douban is collected from~\cite{p34}\footnote{https://github.com/fmonti/mgcnn}. We adopt a 70\%-10\%-20\% train-validate-test division ratio for these datasets in all experiments involved.

{\bf Evaluation Metrics.} To evaluate the accuracy of the missing rating prediction of the tested model, we adopt the root mean square error (RMSE) and mean absolute error (MAE) as the evaluation metrics:
\begin{small}
\begin{align}
R M S E &=\sqrt{\left(\sum_{\substack{\mathbf{x}^{(k)} \in \Gamma \\
\mathbf{m}^{(k)} \in M}}\left(\left(\mathbf{x}^{(k)}-\hat{\mathbf{x}}^{(k)}\right) \odot \mathbf{m}^{(k)}\right)^{2}\right) /\left\|\Gamma\right\|_{0}} 
\end{align}

\begin{align}
M A E &=\left(\sum_{\substack{\mathbf{x}^{(k)} \in \Gamma \\
\mathbf{m}^{(k)} \in M}}\left|\left(\mathbf{x}^{(k)}-\hat{\mathbf{x}}^{(k)}\right) \odot \mathbf{m}^{(k)}\right|_{a b s}\right) /\left\|\Gamma\right\|_{0}
\end{align}
\end{small}

\noindent where $\Gamma$ denotes the testing set and $|\cdot|_{abs}$ denotes the absolute value of a given number.

{\bf Baselines.} The proposed MMA model is compared with 7 state-of-the-art models, including one original model (AutoRec), two Latent factor analysis-based (LFA-based) models (MF and FML), and five deep-learning models (NRR, SparseFC, IGMC, and GLocal-K). Table~\ref{tab:competitor} gives a brief description of these competitors.

\subsection{Performance Comparison}

Table~\ref{tab:comparison} records the prediction accuracy of all models involved in D1 to D5. From Table~\ref{tab:comparison} we can find that MMA achieves the highest prediction accuracy for missing user data prediction than other models.

\begin{table*}[h] 
	\centering
	\setlength{\tabcolsep}{12pt}
	\caption{The comparison of the prediction accuracy of MMA and its competitors, including the loss/tie/win counts, Wilcoxon signed-ranks test, and Friedman test.}
	    \begin{threeparttable}
    	\begin{tabular}{cccccccccc}
    		\midrule
    		{\bf Dataset} & {\bf Metric} & {\bf MF}& {\bf AutoRec} & {\bf NRR} & {\bf SparseFC} & {\bf IGMC} & {\bf FML} & {\bf Glocal-K}& {\bf\begin{tabular}[c]{@{}c@{}} MMA \\ (ours) \end{tabular}} \\
    		\midrule
    		D1 & \begin{tabular}[c]{@{}c@{}} RMSE \\ MAE \end{tabular} & \begin{tabular}[c]{@{}c@{}} 0.857 \\ 0.673 \end{tabular} & \begin{tabular}[c]{@{}c@{}} 0.847 \\ 0.667 \end{tabular} & \begin{tabular}[c]{@{}c@{}} 0.881 \\ 0.691 \end{tabular} & \begin{tabular}[c]{@{}c@{}}  0.839 \\ 0.656 \end{tabular} & \begin{tabular}[c]{@{}c@{}} 0.867 \\ 0.681\end{tabular} & \begin{tabular}[c]{@{}c@{}} 0.849 \\ 0.667 \end{tabular} & \begin{tabular}[c]{@{}c@{}}  0.839 \\  0.655 \end{tabular} & \begin{tabular}[c]{@{}c@{}}0.840 \\ 0.656 \end{tabular} \\ 
    		D2 & \begin{tabular}[c]{@{}c@{}} RMSE \\ MAE \end{tabular} & \begin{tabular}[c]{@{}c@{}} 0.913\\0.719 \end{tabular} & \begin{tabular}[c]{@{}c@{}} 0.897 \\ 0.706 \end{tabular} & \begin{tabular}[c]{@{}c@{}} 0.923\\0.725 \end{tabular} & \begin{tabular}[c]{@{}c@{}} 0.899\\0.706 \end{tabular} & \begin{tabular}[c]{@{}c@{}} 0.915\\0.722 \end{tabular} & \begin{tabular}[c]{@{}c@{}} 0.904\\0.718 \end{tabular} & \begin{tabular}[c]{@{}c@{}} 0.892\\0.697 \end{tabular} & \begin{tabular}[c]{@{}c@{}}  0.889\\ 0.695 \end{tabular}   \\
    		D3 & \begin{tabular}[c]{@{}c@{}} RMSE \\ MAE \end{tabular} & \begin{tabular}[c]{@{}c@{}} 0.757\\0.572 \end{tabular} & \begin{tabular}[c]{@{}c@{}} 0.752 \\ 0.569 \end{tabular} & \begin{tabular}[c]{@{}c@{}} 0.774\\0.583 \end{tabular} & \begin{tabular}[c]{@{}c@{}} 0.749\\0.567 \end{tabular} & \begin{tabular}[c]{@{}c@{}} 0.769\\0.582 \end{tabular} & \begin{tabular}[c]{@{}c@{}} 0.754\\0.573 \end{tabular} & \begin{tabular}[c]{@{}c@{}} 0.756\\0.573 \end{tabular} & \begin{tabular}[c]{@{}c@{}}  0.744\\ 0.561 \end{tabular}   \\
    		D4 & \begin{tabular}[c]{@{}c@{}} RMSE \\ MAE \end{tabular} & \begin{tabular}[c]{@{}c@{}} 1.206\\0.937 \end{tabular} & \begin{tabular}[c]{@{}c@{}} 1.172 \\ 0.900 \end{tabular} & \begin{tabular}[c]{@{}c@{}} 1.227\\0.949 \end{tabular} & \begin{tabular}[c]{@{}c@{}} 1.203\\0.915 \end{tabular} & \begin{tabular}[c]{@{}c@{}} 1.133\\0.848 \end{tabular} & \begin{tabular}[c]{@{}c@{}} 1.176\\0.937 \end{tabular} & \begin{tabular}[c]{@{}c@{}} 1.204\\0.905 \end{tabular} & \begin{tabular}[c]{@{}c@{}} 1.163\\0.879 \end{tabular}   \\
    		D5 & \begin{tabular}[c]{@{}c@{}} RMSE \\ MAE \end{tabular} & \begin{tabular}[c]{@{}c@{}} 0.738\\0.588 \end{tabular} & \begin{tabular}[c]{@{}c@{}} 0.744 \\ 0.588 \end{tabular} & \begin{tabular}[c]{@{}c@{}}  0.726\\ 0.573 \end{tabular} & \begin{tabular}[c]{@{}c@{}} 0.745\\0.587 \end{tabular} & \begin{tabular}[c]{@{}c@{}} 0.751\\0.594 \end{tabular} & \begin{tabular}[c]{@{}c@{}} 0.762\\0.598 \end{tabular} & \begin{tabular}[c]{@{}c@{}} 0.737\\0.580 \end{tabular} & \begin{tabular}[c]{@{}c@{}} 0.740\\0.581 \end{tabular}   \\
    		\bottomrule
    	\end{tabular}
    \end{threeparttable}
	\label{tab:comparison}
\end{table*}

%% file: Conclusion.tex
\section{Conclusion}
\label{sec:conclusion}
This paper proposes a multi-metric Autoencoder model to accurately predict missing user behavior data. Its essential idea is two-fold. Firstly, deploy different $L_p$-norms as loss function and regularization to form four base models with different metric. Then combine these base models with the adaptive weighting strategy. Experiments on five real-world datasets prove that MMA has remarkably higher accuracy in predicting missing user behavior data.